**Institute of computer systems**
**Department of physics**

**Research report**

**Peculiarities of image acquirement and analysis in AFM electrical modes**

Olga V. Sviridova (Ольга Валентинівна Свірідова, Ольга Валентиновна Свиридова)
[1] *Odessa National Polytechnic University, Shevchenko Ave. 1, 65044 Odessa, Ukraine,*
*e-mail: sviridova@onu.edu.ua*

**Abstract:** The paper describes the peculiarities of acquirement and interpretation of images of current distribution through the sample surface when operating in I-AFM mode. It shows that I-AFM and SCM modes can be successfully used only for small scanning fields (no more than 5x5 μm²), since during the scanning process the continuous change in the area of the probe tip and, therefore, in the contact area between the probe and the sample surface is observed because of the abrasion of the tip. At the same time electrical modes of AFM could not be recommended for the investigation of nano objects, because there appear a number of difficulties in results interpretation, caused by the big curvature radius of the probe tip and, therefore, by the big surface area of electrical contact.

The paper also demonstrates the peculiarities of acquirement and interpretation of CVCs for individual points on the sample surface in I/V Spectroscopy mode. It is shown that it is practically impossible to use I/V Spectroscopy mode for additional investigation of the surface by acquiring of CVCs in the points of interest (where heterogeneities in topography or current through the surface are observed) on the sample surface in I-AFM mode, because of the big temperature drift and hysteresis of piezoceramics. Recommendations for improving the possibilities of the method are given in the paper.

**Introduction**

In recent years, scanning probe microscopes (SPMs) are becoming more common. The most common type of SPM is AFM, which as a rule operates in three basic modes: non-contact, contact and semi-contact (tapping mode). In addition to these modes manufacturers of AFM widely develop and introduce various additional modes such as current AFM (I-AFM), I/V Spectroscopy, scanning capacitance microscopy (SCM), scanning spreading resistance microscopy (SSRM), electrostatic force microscopy (EFM), etc. [1-3]. Unfortunately, in the accompanying user manuals, manufacturers of microscope report only on how to switch to the required mode and how the specialized software works in this mode. Practically nothing is reported on the peculiarities of work in a particular mode, on sample preparation, on artifacts, as well as on the correct understanding and analysis of measurement results. In a number of publications [4-6], which use measurements obtained by electrical modes of SPM, there is also no information about the methodology and peculiarities of receiving presented results – it is only reported on the used mode of SPM. The main reason for this is that the SPMs are constantly upgraded (sometimes faster than instructions to them are written). Thus, both in publications and user manuals a detailed methodology of measurement is constantly overlooked. This can often lead to an incorrect understanding and, as a result, incorrect interpretation of the data presented in the papers, as well as it complicates the possibility of repetition of experiments, described in the papers, for the purpose of their finalization and modification. Meanwhile, an inexperienced user (or even a user, experienced in topography receiving modes) has to spend a lot of time in order to realize himself the possibilities of the device and the used method, and to learn how to receive and interpret the results correctly. This is connected to the fact that when operating in the above modes, there arise a number of features that never, under any circumstances arise when operating in topography modes.

Therefore, the presented paper provides important technical moments that researchers should take into account when operating in electrical modes of AFM. It demonstrates their minuses and pluses on the example of SPM investigation of Schottky contacts in ZnO-based Schottky diodes.

## Equipment
All results, presented in the paper, were obtained on AFM XE-150, manufactured by Park Systems (South Korea), in the laboratory of Semiconductor Physics group, Universität Leipzig. A detailed description of operating principles and possibilities of I-AFM, I/V Spectroscopy, SSRM and SCM modes is given on the web-site of Park Systems [7]. Probes NSS18Ti-Pt and NSS36Ti-Pt [8] were used for the measurements.

## Discussion
### 1. I-AFM mode
#### The idea of the method
In I-AFM mode of XE-150 by applying a positive or negative potential to the sample, one can perform simultaneous acquisition of the surface topography and the image of current distribution through the sample surface. Thus the probe is always grounded. In order to apply a potential to the sample, it is necessary to fix the sample motionless on a metal substrate and to establish an ohmic contact between the sample and the substrate [7]. Thereafter, the metal substrate with a sample is set on a special magnetic table inside of SPM, which is electrically connected to the device. By varying the potential difference one can increase or decrease current through the sample. Current amplification is performed by using one of the 3 amplifiers (1 built-in and 2 external). Knowing the amount of current flowing through a given "point" on the sample surface, the applied voltage and the contact area of the probe tip, this method on principle allows the calculation of electrical resistance or current density at the point of measurement.

#### The formation of ohmic contact between the sample and the metal substrate
The studied samples were ZnO-based diodes, with the following structure: a-sapphire substrate/back ohmic contact based on ZnO:Ga/epitaxial layer of ZnO/metal oxide ($PtO_x$, $PdO_x$) – Schottky contact realized without any pretreatment by reactive DC sputtering through a lithographic mask (see Fig. 1) [9].

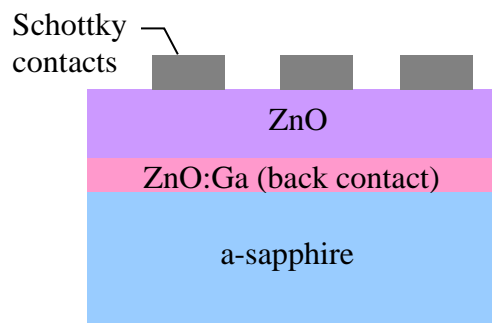

Fig. 1. Schematic representation of cross section of studied structures.

In order to apply a voltage to the back contact, the gold layer (~ 40 nm) is traditionally sputtered in an atmosphere of argon at the end face of the sample (see Fig. 2). Gold forms an ohmic contact to ZnO:Ga.

In order to make measurements in I-AFM mode, one must first of all establish an ohmic contact between the metal substrate and the sample. As a rule, it is sufficient to apply a small amount of two-component conductive silver paste on a metal substrate. After this one should position the end face of the sample at the site of application of silver paste and lightly press the sample to the substrate. Thus a part of silver paste penetrates to the reverse side of the sample (and fixes the sample on a metal substrate), a part remains on the end face, and a part penetrates to the sample surface (see Fig. 3). The thickness of silver paste layer varies in different parts of the sample surface (on the end face and on the back surface) and depends on the amount of paste, initially applied by the user on a metal substrate, and on pressing force of the sample to the substrate. As a rule the thickness ranges from 0.1 mm on the back surface of the sample to 5 mm at the end face.

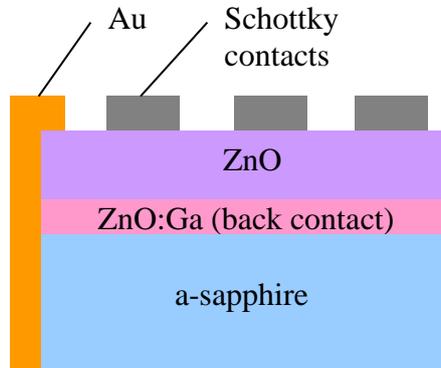

Fig. 2, a. Schematic representation of cross section of Schottky diodes with Au sputtered at the edge in order to contact the back contact.

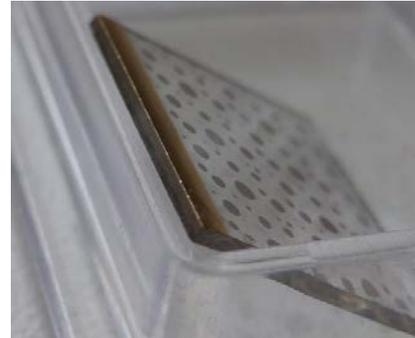

Fig. 2, b. Image of Schottky diodes with Au sputtered at the edge in order to contact the back contact.

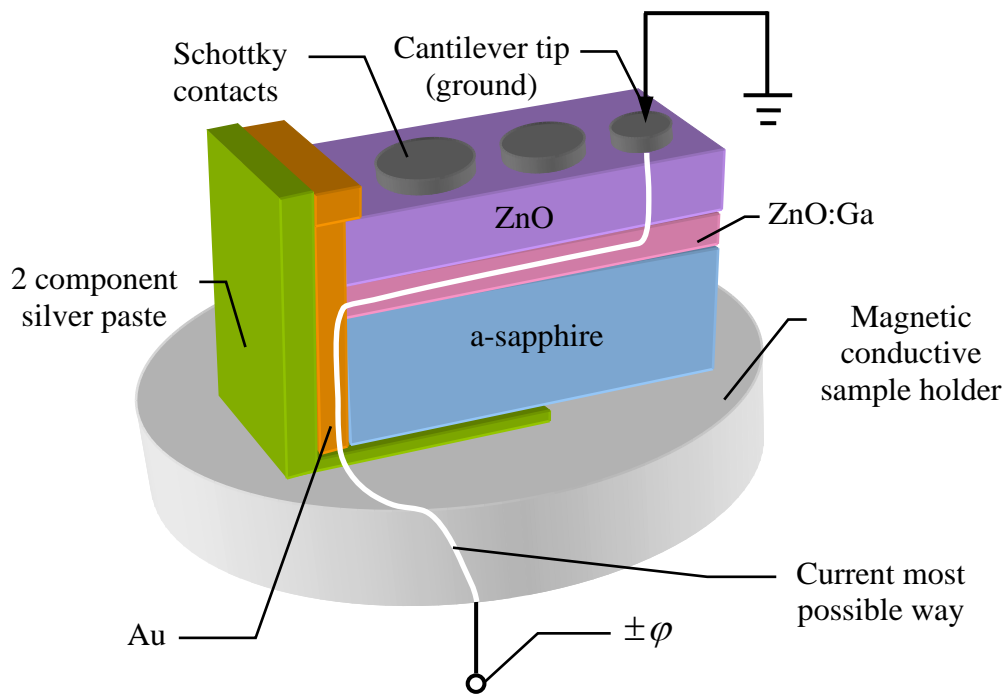

Fig. 3, a. Schematic representation of Schottky diodes mounted on a magnetic sample holder with 2 component silver paste.

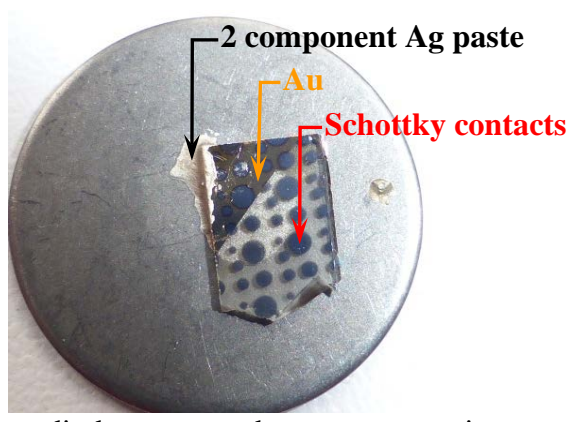

Fig. 3, b. Image of Schottky diodes mounted on a magnetic sample holder with 2

After hardening of silver paste, electrical contact is formed between gold, sputtered at the end face, and silver paste as well as between the metal substrate and the sample. Thus, applying a potential to the metal substrate, we at the same time apply the potential to the silver paste, and through it – to the gold layer sputtered on the end face. Using an Agilent 4155C Semiconductor Parameter Analyzer connected to a SÜSS Wafer Prober PA200 HS, we applied a potential difference between the metal substrate and the Schottky contact on the surface of the diode. The presence of CVCs, typical for Schottky diodes, indicated that electrical contact between the diode and the substrate is operating normally. However, when installing the sample in AFM and trying to get a current image, we failed to detect any electrical signal throughout the operating range of current amplifiers that we used. The increase of applied voltage was also inconclusive. Repeat contact quality control using an Agilent 4155C Semiconductor Parameter Analyzer connected to a SÜSS Wafer Prober PA200 HS showed the presence of current through the sample. It seemed that the problem is in the ground, carried out through the probe – either the wire, that contacts the platform with cantilevers and resonator, was damaged, or the probe had a defective metal coating. A multiple replacement of cantilevers and the change of probes of one and the same cantilever (C, A, B) did not give any result. At the same time, repeat sputtering of gold at the end face of the sample (see Fig. 4) gave a positive result.

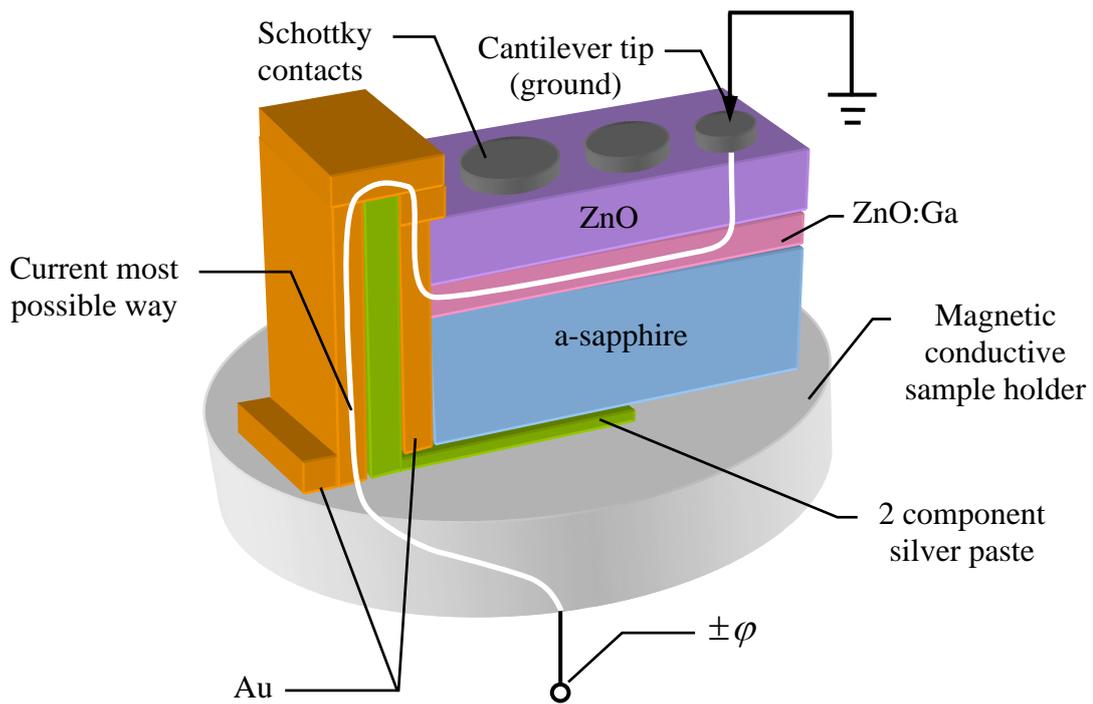

Fig. 4, a. Schematic representation of Schottky diodes mounted on a magnetic sample holder with 2 component silver paste after repeat Au sputtering.

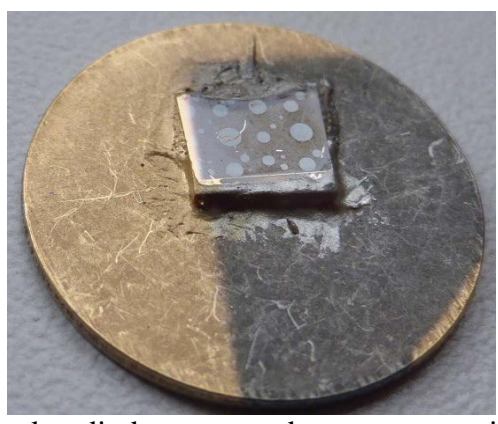

As it turned out, when using the probe tip as a contact, the current through the sample is so small that the voltage drop, caused by current flow through the layer of silver paste, leads to impossibility of signal detection. In the case of Wafer Prober the area of electrical contact greatly exceeded the area of electrical contact of the probe tip, therefore Wafer Prober showed the presence of current through the sample.

Thus, as a result of preliminary studies, for a successful work in I-AFM mode we suggest to mount the samples to the metal sample holder in the following way: on the end face of the diode a thick layer of gold (~ 100 nm) is sputtered in the argon atmosphere, then up to 30% (for strong fixation) of the length of the end face is fixed to the sample holder using the silver paste. After hardening of silver paste, gold (~ 100 nm) is repeatedly sputtered in an argon atmosphere on the entire end face of the sample (see Fig. 5).

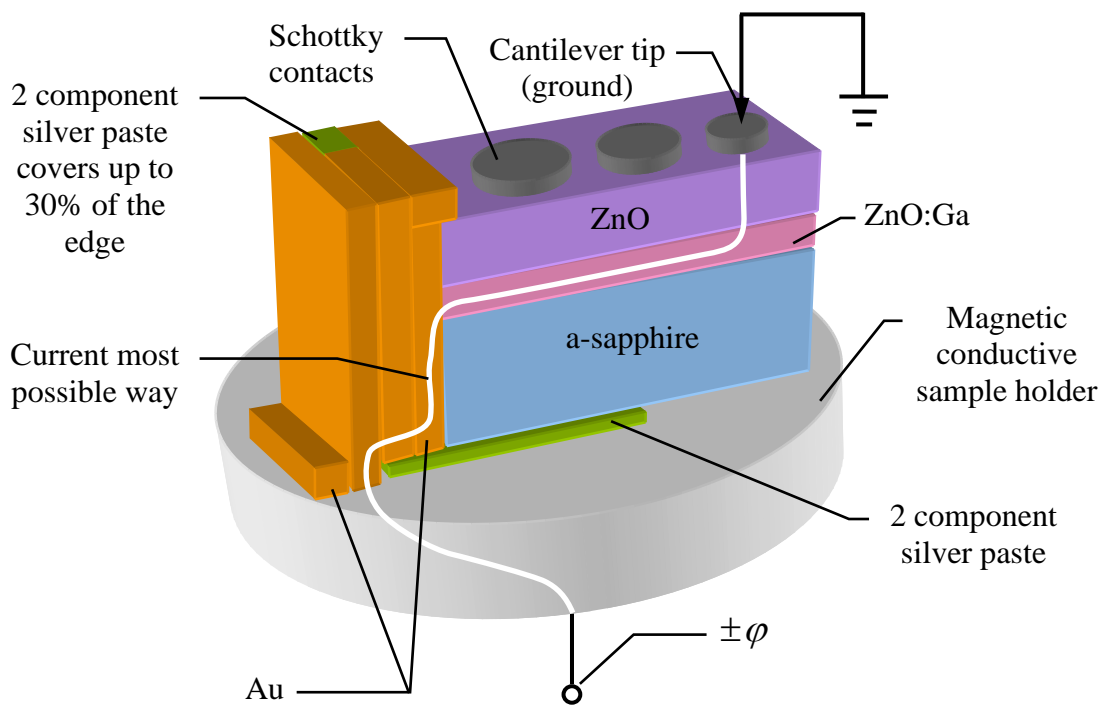

Fig. 5, a. Schematic representation of recommended fixation of Schottky diodes with 2 component silver paste and repeat Au sputtering.

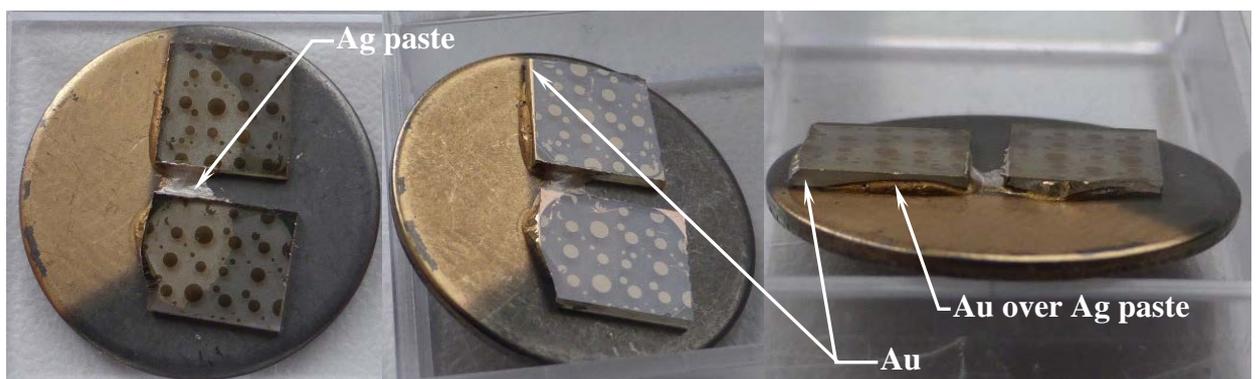

Fig. 5, b. Images of recommended fixation of Schottky diodes with 2 component silver paste and repeat Au sputtering.

Thus, if in the initial experiments when preparing the sample a series connection of metal substrate, silver paste, gold and ZnO:Ga was formed, and the current necessarily flowed through the silver paste, then in the end it flows along the surface of gold, coating the silver paste, and through those part of the end face of the sample surface, which is coated only with gold. This allows to avoid the voltage drop, when current flows through the porous ohmic layer of silver paste.

**Change in the parameters of probe during the scanning process**

Operating in I-AFM mode is always an operation in contact mode with probes, the resulting tip curvature radius with the coating of which is less than 40.0 nm [8]. Therefore, one should not initially expect to receive a high-quality surface topography image.

When one operates in a topography mode, whether it is a contact, non-contact, or semi-contact mode, the state of the probe can be judged primarily on the quality of image and on the forward and backward trace line in the corresponding trace control window. When operating in I-AFM mode it is not enough.

I-AFM mode is intended primarily for electrical measurements. Topography is secondary in this mode. For operation in I-AFM mode it is important that the parameters of electrical system (when receiving a current image) remain constant during the whole scanning process. These parameters include: applied voltage, used signal gain, and the area of electrical contact. Only when performing all these conditions we can say that the change in the current, flowing through a given point of the sample surface, is caused by heterogeneity of the crystal structure of the sample, or by changes in its chemical composition. If the constancy of the first two parameters the user can control in the settings, then with the constancy of the last parameter – the contact area, there is a certain problem, connected with the fact, that the contact area of the sample with the probe varies from one scanning point to another.

It is, first of all, connected with the inhomogeneity of the relief of the sample surface. In contrast to the flat areas of the sample surface, when the contact between the sample and the probe takes place usually on the probe tip, for the surface areas containing a large number of narrow indentations or steep uplands, the electrical contact between the probe and the surface is mainly due to the lateral surface of the probe (as a blunt probe mechanically can not penetrate into narrow gaps, and when scanning any uplands, current often flows both through the tip and the lateral surface of the probe, that is adjacent to the sample surface), Fig. 6.

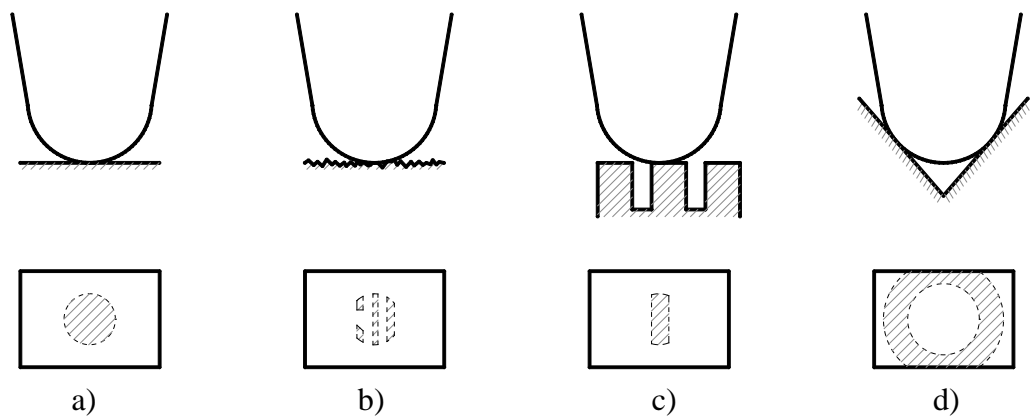

Fig. 6. Models of probe interaction with various surface relief: a) – flat surface; b) – a surface with the developed relief; c) – a surface with projections or narrow grooves; d) – a surface with an angular groove.

Besides the sample surface, the area of electrical contact and, hence, the magnitude of the electric current is affected by the state of the conductive metal coating, deposited on the probe tip. Thus, for a homogeneous sample with low surface roughness and a new probe, electrical contact is formed between the probe tip and the sample surface. During the scanning process there is a gradual erasure of metal coating of the tip and an increase in the contact area, resulting in an increase in the magnitude of detected current. With time this coating is erased completely, but conductive layer still remains on the perimeter of the probe tip (see Fig. 7). Therefore, for erased probes the presence of current will be observed on heterogeneous areas of sample surface, on which there is a contact between the lateral surface of the probe and the sample material (see Fig. 8 – 10).

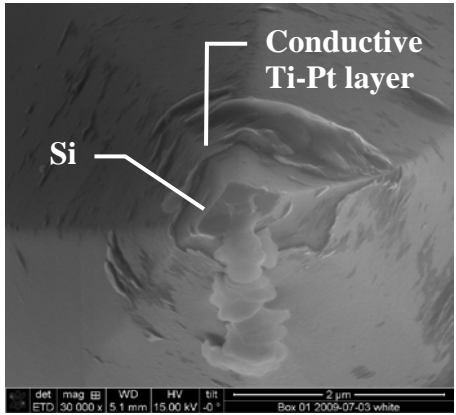

Fig. 7. SEM image of damaged NSC 18 Ti-Pt cantilever tip, taken after it was impossible to register any current.

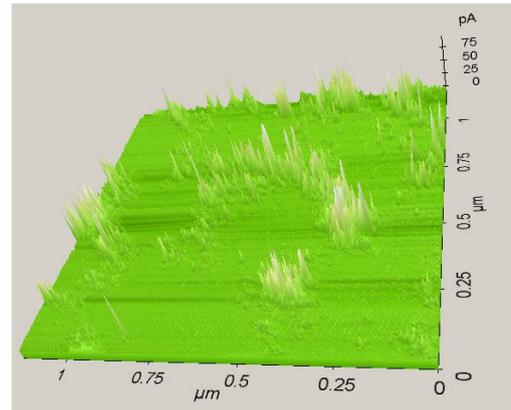

Fig. 8. 3D I-AFM image of Schottky contact obtained with NSC 18 Ti-Pt cantilever (see Fig. 7), when the tip was a little bit less damaged.

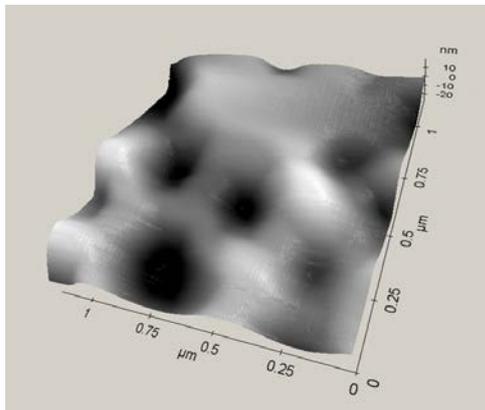

Fig. 9. 3D C-AFM image of Schottky contact obtained with NSC 18 Ti-Pt cantilever (see Fig. 7), when the tip was a little bit less damaged.

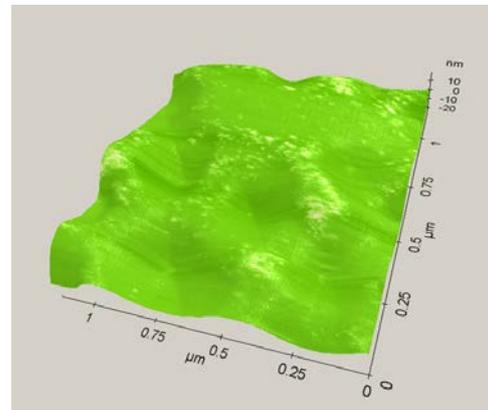

Fig. 10. Overlap of 3D C-AFM image (see Fig. 9) and 2D I-AFM image of Schottky contact obtained with NSC 18 Ti-Pt cantilever (see Fig. 7), when the tip was a little bit less damaged. Color corresponds to the values of current (see Fig. 8).

Since chips with cantilevers are very fragile, and assembly and disassembly of chips, both in AFM and SEM is always connected with the risk of damaging them, we had no opportunity to get SEI images of probes on completion of each scanning stage. Deterioration of the conductive coating of the probe tip was indicated by current trace lines in the corresponding trace control window. SEI image, shown in Fig. 7, was obtained after no current could be registered using this probe. Therefore, in comments to Fig. 8 – 10 it is pointed out that at the time of image acquisition the surface of the probe was less worn out.

The main problem when operating in I-AFM mode is that the user has no idea about the state of the conductive coating on the probe tip during the whole scanning process. It would seem that to the presence of a conductive coating on the probe tip should correspond a trace line in the current trace control window, which has a certain heterogeneity (caused by inhomogeneity of the surface) over the entire length, that corresponds to the permanent availability of the contact. In the ideal case forward and backward current trace lines should overlap (see Fig. 11). However, most often it happens that the topography image is perfect, forward and backward topography trace lines optimally coincide, there is no noise, the error signal is minimal, but forward and backward current trace lines match rather arbitrary: for example, signal increase or decay are observed both on forward and on backward scan trace lines, but with a significant shift; the sizes of current maxima and minima may differ both in absolute value and in width (see Fig. 12). The comparison of topography images of the sample, obtained for the fast scan direction from left to right (forward scan direction) and from right to left (backward scan direction), shows that they match, while the comparison of the corresponding current images shows, that there is a shift. There arises a question: what should we trust?

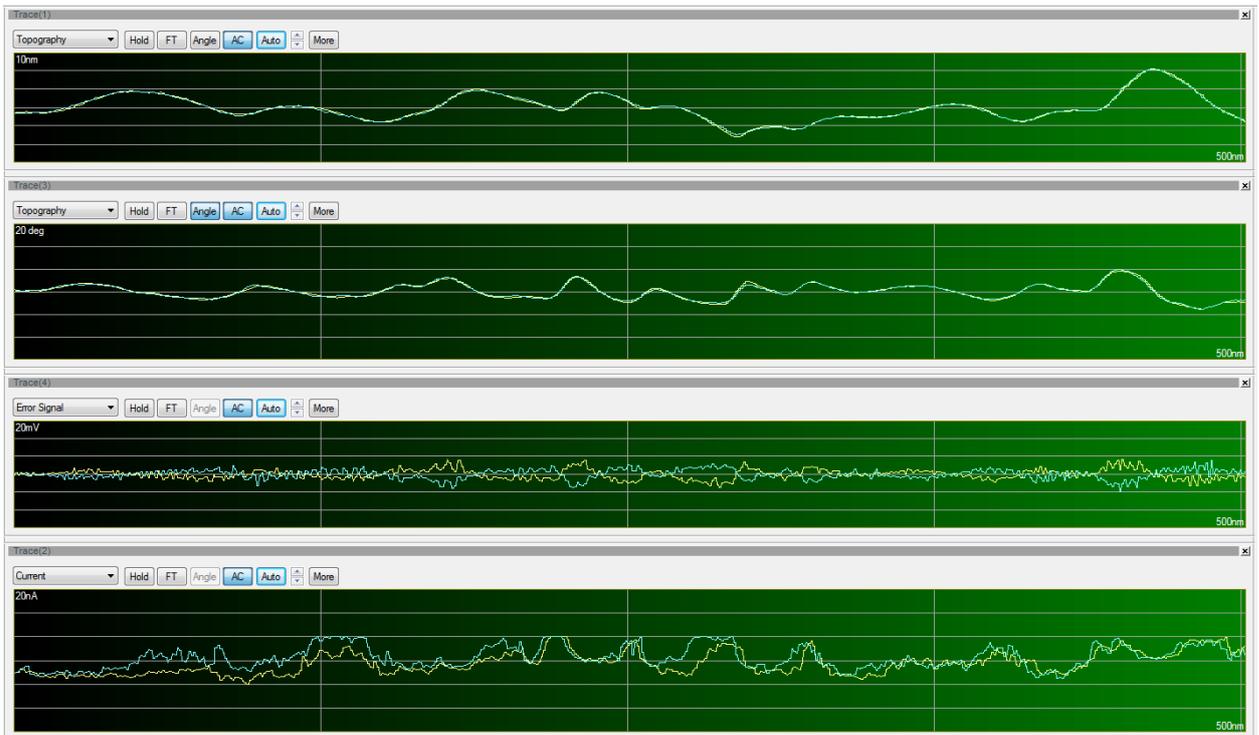

Fig. 11. Screenshots of Topography, Topography Angle, Error Signal and Current trace lines, depicted at the monitor of AFM PC. Yellow trace line corresponds to forward scan direction. Blue trace line corresponds to backward scan direction. From the comparison of trace lines, it is clear that the corresponding to them topography and current images are identical. This is the ideal case.

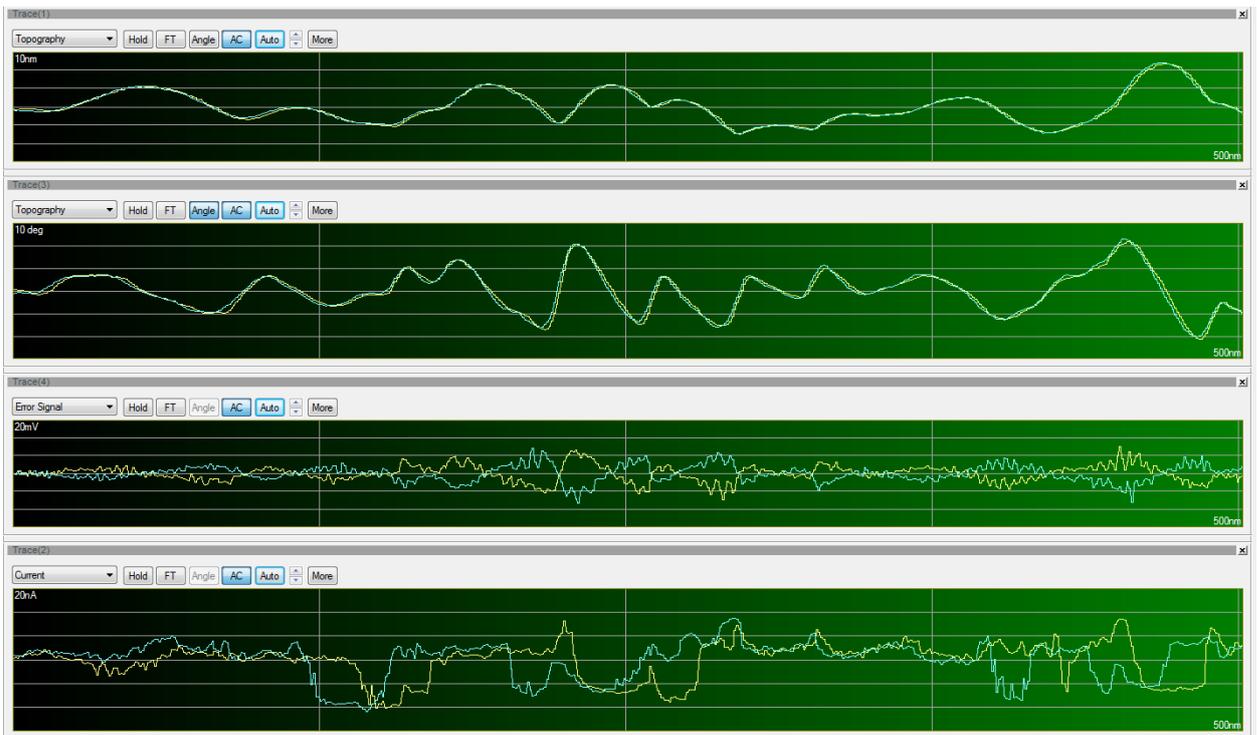

Fig. 12. Screenshots of Topography, Topography Angle, Error Signal and Current trace lines, depicted at the monitor of AFM PC. Yellow trace line corresponds to forward scan direction. Blue trace line corresponds to backward scan direction. From the comparison of topography trace lines it is clear that the corresponding to them images are identical, while the comparison of current trace lines shows the difference (see shift of forward and backward trace lines at the bottom image).

For the user, experienced in topography imaging, the difference in the images acquired in forward and backward scan directions means that scanning settings such as scanning frequency, set point, proportional gain (P-gain), integral gain (I-gain), etc. were selected incorrectly. But what should the user do when topography images are identical and current – different (see Fig. 13)?

Here it is necessary to consider the following: the topography image is obtained by detecting and subsequent processing of the signal (laser beam), reflected from the outer side of cantilever and reaching the four-sector photodetector. The current image is built on a completely different principle. When building a current image, the SPM processes current signal, that comes either from a built-in internal or external current amplifier. The amplifier increases to the required order the current, that flows between the conductive surface of the probe, contacting the sample, and the sample itself. Thus, the amount of current, detected at each scanning point, is determined by the contact area of the probe with the sample surface. In principle, a current image can be obtained when the laser and the photodetector are switched off – there is no need in them for that (Actually, in I/V Spectroscopy mode it is possible to switch off the laser, when receiving the CVCs, in order to avoid photocurrent in semiconductors, photosensitive to the wavelength of the used laser.). Another thing is that the current image itself does not bear any specific information, if you can not compare it with the topography of the surface. In addition to the influence of surface relief of the sample and the design features of the probe (asymmetry), one possible explanation for the observed in Fig. 13 signal shifts, may be the presence of liquid or dust on the sample surface [10]. It is no secret that, for obtaining a high-quality image in contact mode, it is better to use continuous scanning, when at the end of the last line in the selected scanning field, the scanner without stopping changes the direction of slow scanning and starts to scan backwards. As a rule, after one and the same area was scanned twice, the third scan gives several times better image quality. The same principle should be valid for the current mode.

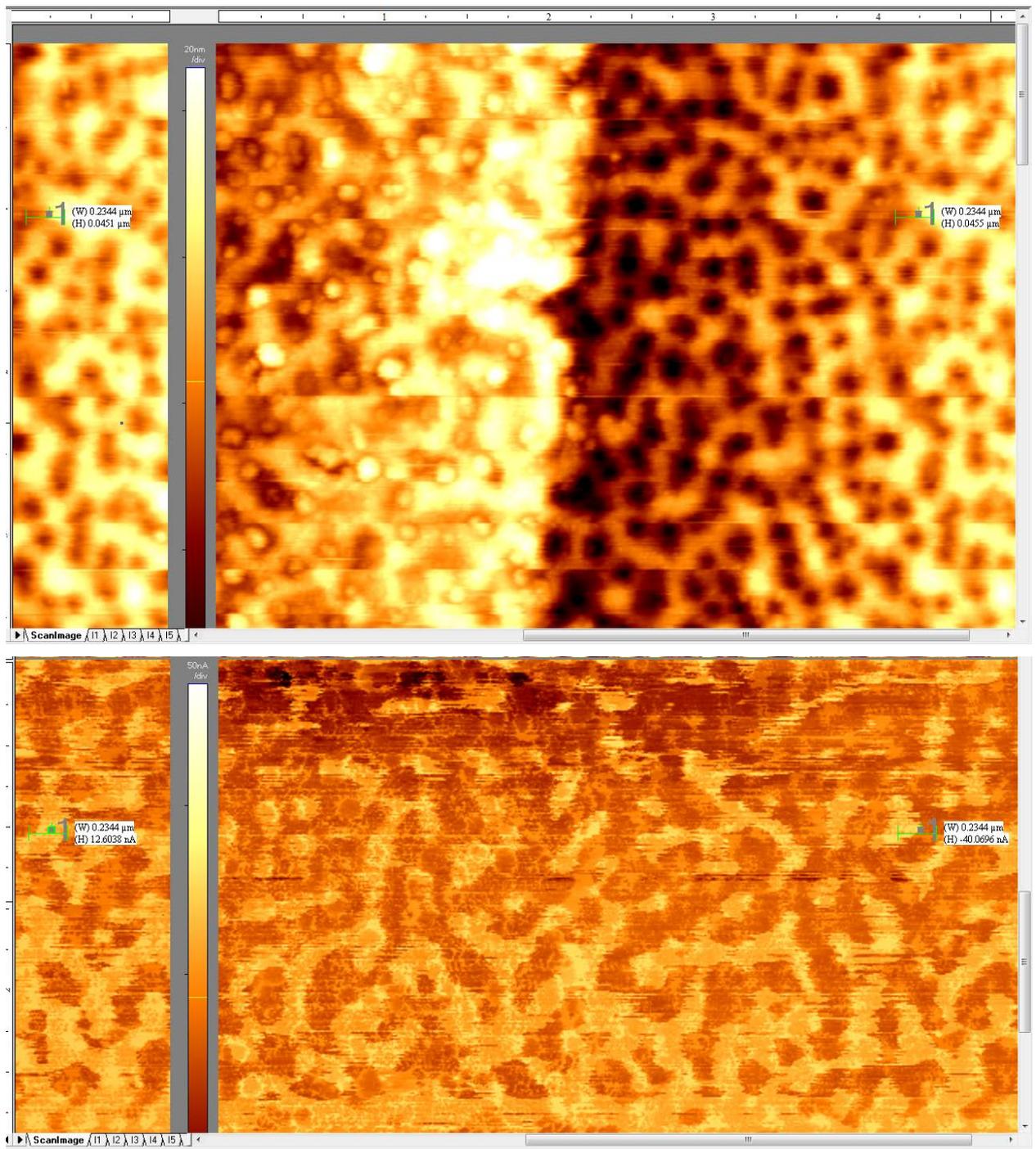

Fig. 13. Screenshots of topography (top) and current (bottom) images depicted at the monitor of AFM PC. Both images belong to one and the same scanning and were taken simultaneously. They correspond to the trace lines depicted at Fig. 11. At the left part of images one can see a part of forward scan direction image. At the right part of images one can see an image, corresponding to the backward scan direction. From the comparison of topography images it is clear that they are identical, while the comparison of current images shows the difference (see green marker located in one and the same point of the sample).

## Continuous (repeat) scanning

Continuous, or as it is called by manufacturers of XE-150, repeat scanning, certainly should lead to the clearing of the sample surface from the condensed liquid and dust, which are always present on the surface. However, in contrast to the contact mode of AFM, when continuous scanning at properly chosen probe and settings does not lead to a significant change in the probe tip and has positive effect on the quality of measurements, when using conductive probes continuous scanning cannot be recommended.

This is due to the fact that NSC18Ti-Pt and NSC36Ti-Pt probes quite rapidly (in few hours, and in some cases, after several scans about $5 \times 5$ mm$^2$) become worn. The technological feature of manufacture of such probes is that a thin conductive coating is deposited on a previously manufactured dielectric probe, the surface of which has certain inhomogeneities and porosity. Surface inhomogeneity of probe reduce the adhesive strength of the coating [11]. Furthermore, the process of current flow, which leads to the probe heating, may also serve as one of the reasons of the exfoliation of conductive coating (see Fig. 7). Taking into account the difference in the thermal expansion coefficients of silicon, titanium and platinum, heating and mechanical loads during scanning with subsequent cooling at the end of scanning, may lead to the exfoliation of metals, deposited on silicon.

In view of the above, the option of continuous scanning cannot be recommended for the I-AFM, as the abrasion of the probe occurs very quickly.

This means that it is not possible to eliminate the difference in the current images, corresponding to forward and backward scan directions, and caused by liquid or dust (see Fig. 14 – 15).

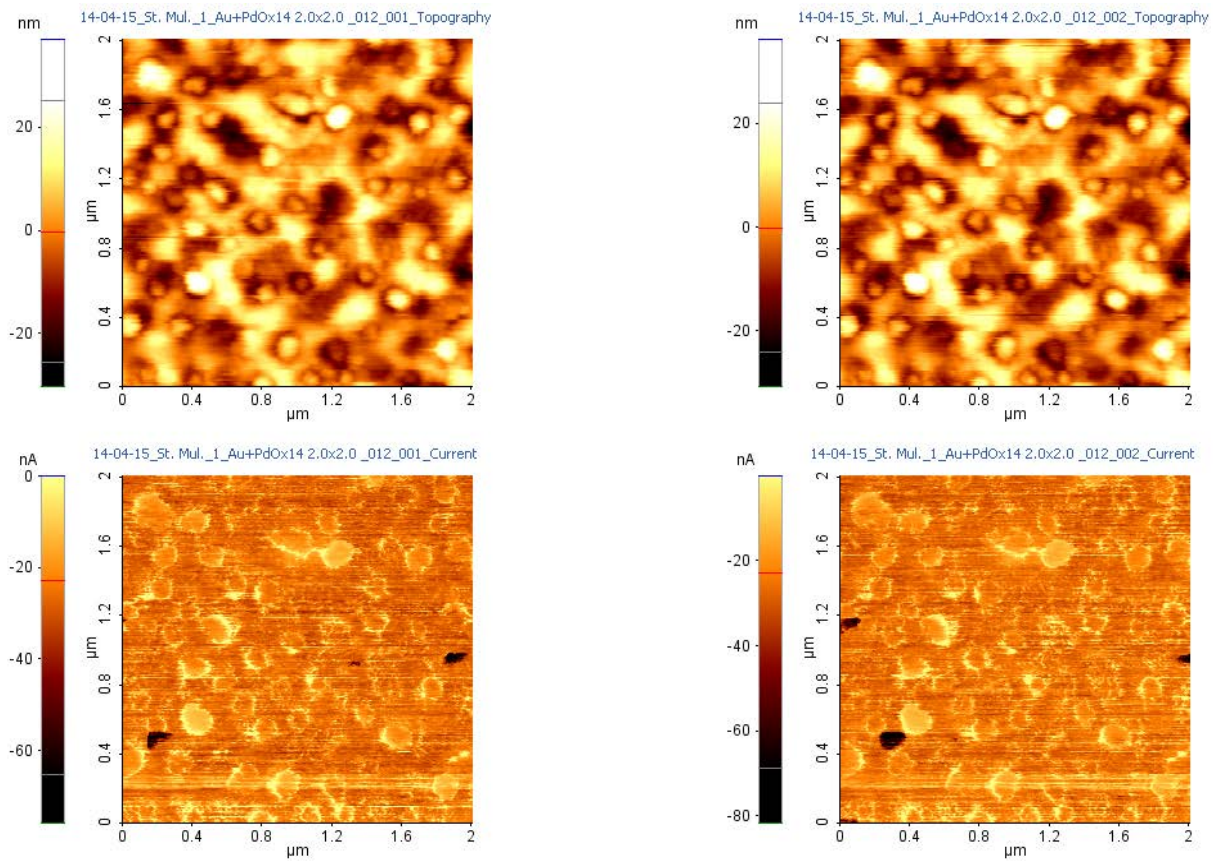

Fig. 14. Topography (top) and corresponding to them Current (bottom) images.
001 corresponds to forward scan direction;
002 corresponds to backward scan direction.
One can see that Topography images are identical, while Current images slightly differ.

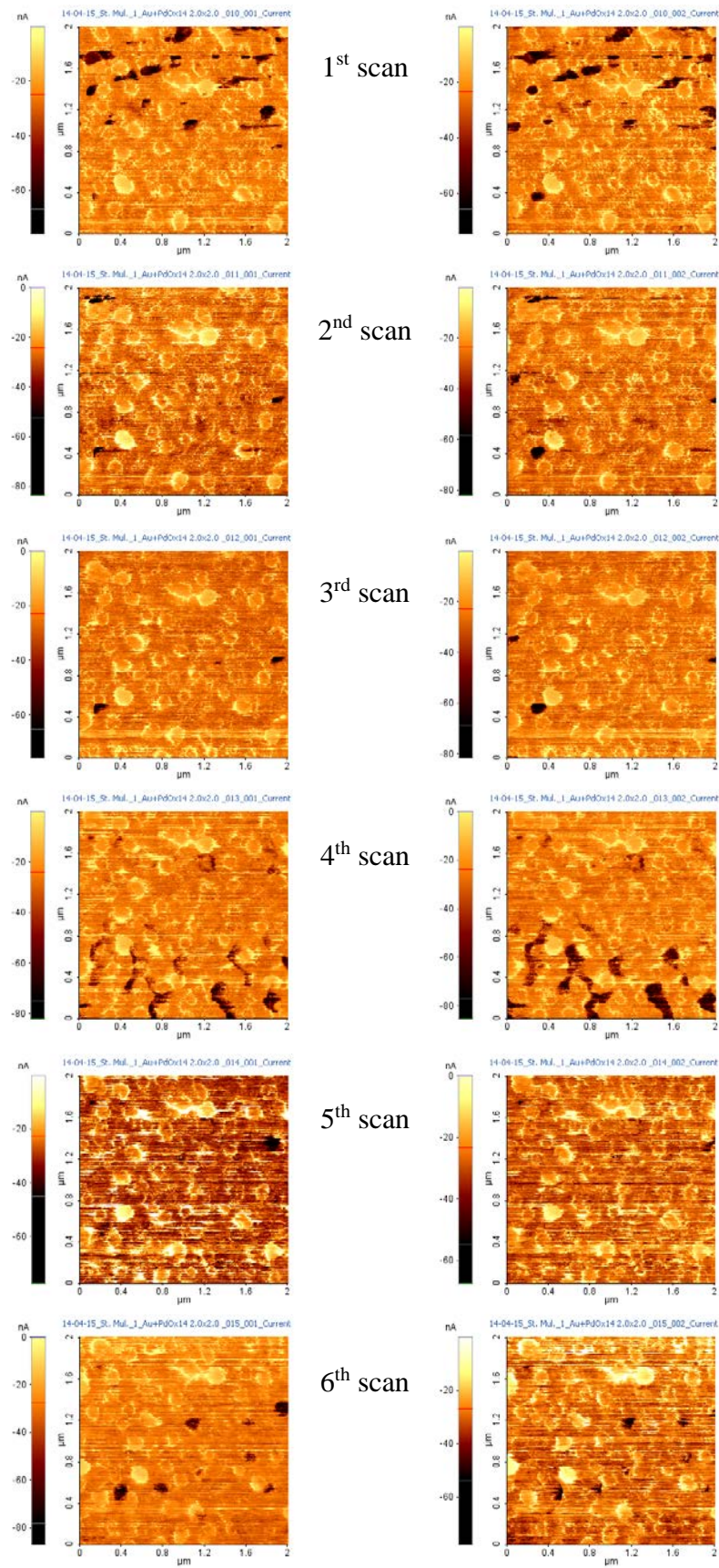

Fig. 15. Current images of one and the same surface area, obtained using the repeat scan option. Left column corresponds to forward scan direction (001_Current), right column corresponds to backward scan direction (002_Current). One can see that for the $4^{th}-6^{th}$ scans the current image is less sharp, than for three previous scans, and plenty of artifacts appear, i.e. the scanning process spoils the probe tip rather fast.

**The study of the feasibility of using cantilevers with three probes.**

The matter is that the cantilever replacement process when operating in I-AFM mode is somewhat more complicated, than when operating in the topography mode: the cantilever chip is connected to the Head Extension Module (HEM) for I-AFM via a fragile wire. In the process of mounting the Chip Carrier with cantilever chip to the Probe Hand there is a great risk of damaging this wire, resulting in unsuitability of the cantilever as a whole. Therefore, frequent replacement of cantilever chips mounted on the Chip Carriers is not only time consuming, but is also connected with the risk of damaging the Chip Carrier with conductive cantilever during assembly.

It would seem that using chips with three cantilevers of different lengths (C, A, B) Fig. 16, may serve a good solution to prevent rapid wear of probes and, therefore, frequent replacement of cantilever chips. Such chips with three conductive cantilevers, for example NSC36Ti-Pt, are produced both for I-AFM, EFM and SCM modes. One and the same cantilever chips are mounted on different Chip Carriers and are supplied with different types of wires, in accordance with the design features of AFM when operating in each mode.

However, the use of such cantilevers is highly questionable. Chips with three cantilevers were initially designed for operation in topography mode [12]. All three cantilevers: A, B and C have different lengths. When using all the three cantilevers, scanning always starts with the longest one. As the longest cantilever worns down, the operator moves to the cantilever of medium length, and then to the shortest cantilever.

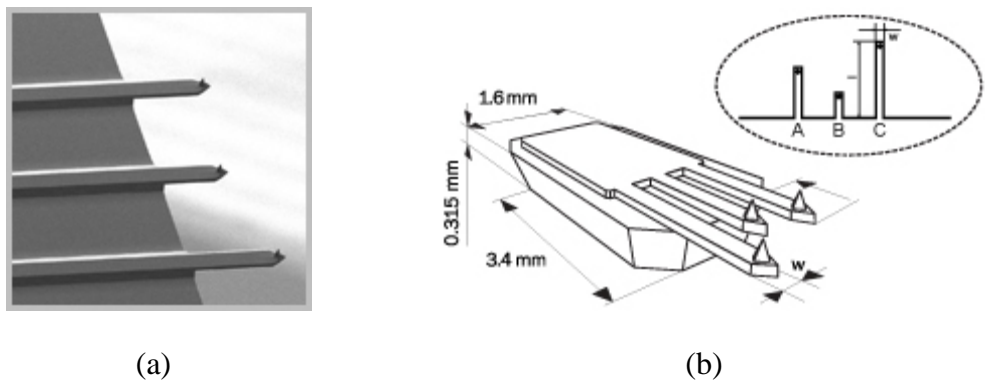

(a)                                                        (b)

Fig. 16. SEM image (a) and schematic drawing (b) of 3 cantilevers on the chip (NSC35/36 series) [2].

One should not forget about the difference in the mechanism of imaging when scanning in topography mode and electrical modes. When operating in the topography mode, the image is constructed on the basis of the changing position of laser beam, reflected from the back side of cantilever and reaching the photodetector. Since the AFM uses only one laser beam, which is focused on one of three cantilevers, when operating in contact mode of AFM using a chip with three cantilevers (3-lever series), one should keep in mind that only the movement of those cantilever, on which the laser beam is focused, is used for imaging. As the use of cantilevers begins with the longest one, it is considered that in the process of scanning the other two (shorter) cantilevers do not touch the surface. For flat, clean and smooth samples with roughness of the order of several nanometers, this approach can be considered as fair. After the longest cantilever is worn out, the cantilever of medium length is used, and then – the shortest cantilever (Fig. 16). However, most of the samples are not ideally clean, plane-parallel and mirror-smooth. Therefore, actually all three cantilevers somehow touch the surface during the scanning process: it is always touched by the longest cantilever, on which the laser beam is focused, and periodically by two shorter ones. Sometimes the scanning process using the longest cantilever is going well: the cantilever scans for a long time without any problems, but when the probe is worn out and it is necessary to switch to a new cantilever, it is found out that only one or even no other cantilevers are left (they were snapped off during the scanning process), or the cantilevers are present, but one of them (or both) have damaged probes. And sometimes, all is well with shorter cantilevers. Anyway, the fact that shorter cantilevers can periodically touch the surface or even scan it, does not affect the resulting topography image, obtained with the currently used cantilever, since the image is constructed on the basis of signal coming from one cantilever and the movements of others do not matter.

In chips with three conductive cantilevers the sequence of cantilever arrangement along the chip width from left to right may differ from their sequence in chips, designed for contact AFM, but the principle of sequence of cantilever selection from the longest to the shortest one remains the same: after the abrasion of the longest cantilever (C), it is necessary to use the cantilever of average length (A), and then the shortest cantilever (B) (see Fig. 16).

But, in contrast to cantilevers, used in contact mode, all conductive cantilevers that touch the sample surface contribute to the signal that is transmitted to the current amplifier (see Fig. 17). I.e. it does not matter which of the cantilevers scans the surface at the moment: whether one, two, or three cantilevers touch the surface, the current will flow through everything that has electrical contact with the surface. Even if the probes of the longest and medium cantilevers are abraded and the user scans with the shortest cantilever, one can not exclude the possibility that a current will also flow through the two worn out cantilevers. Firstly, if the shortest cantilever is in work, there will exist a mechanical contact of the probe of this cantilever with the surface, and this automatically means that the long and medium cantilevers will also have a contact with the surface (unless, of course, they did not snap off, what is also possible). And though their probes have erased conductive coating on the tip, that prevents their use in obtaining the current image, the conductive coating still remains on the lateral surface of worn out probes, as well as on the perimeter of worn out tip. This means that in the case of large inhomogeneities, or slope of the sample surface, some part of the lateral surface of worn out probes and conductive coating on the perimeter of their tip will pass an electrical current. Thus, when using a chip with more than one cantilever, never and under no circumstances one can be sure that detected signal, and hence the registered heterogeneities of current at the point of cantilever location are connected with structural heterogeneities of the studied sample in this point, since one can not be sure that the electrical signal comes only from this point of the sample surface.

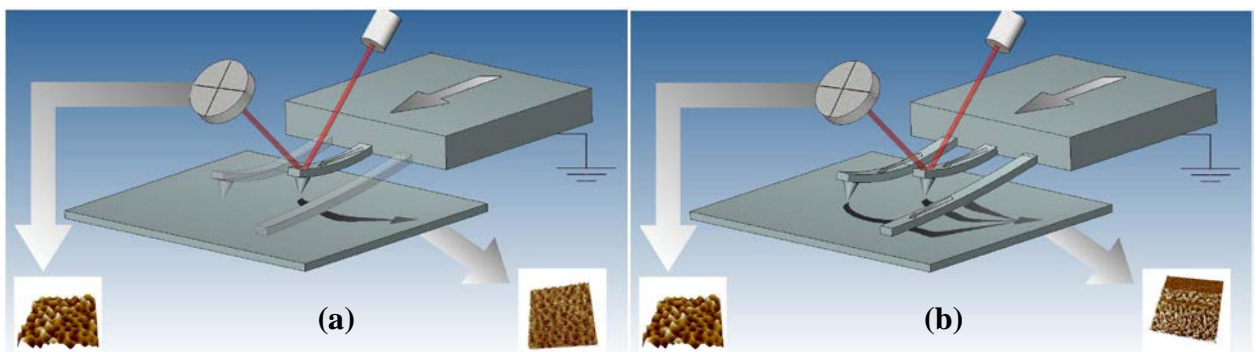

Fig. 17. Schematic representation of difference in imaging with 1-lever (a) and 3-lever (b) cantilevers in electrical modes of AFM.

In view of the foregoing, one can conclude that chips with three cantilevers may be used in only one case: in the study of super smooth (surface roughness about several Å), plane-parallel and clean surfaces and with the proviso that the worn out cantilevers will be mechanically broken off (with tweezers) in order not to cause harmful interference in the detected signal.

Chips with three cantilevers, inter alia, are produced in order among three cantilevers of different lengths, and hence of different values of spring constant, one could select the desired spring constant. Thus it is supposed that two remaining cantilevers must be broken off. The quality of the topography and current images when operating in air depends less on the spring constant as on the temperature drift of piezoelectric ceramics and the quality of the probe tip. Therefore, from a practical point of view, these 3-lever series cantilevers are useless for electrical measurements: if not to delete unnecessary cantilevers, the reliability of the results will be very dubious.

The result of the foregoing is the following: it is not recommended to use chips with more than one cantilever for all electrophysical modes (I-AFM, I/V Spectroscopy, SCM, SSRM, EFM, etc.) of AFM.

## Interpretation of results

Apart from differences in the principle of topography and current imaging, there is a fundamental difference in the correct understanding of obtained results.

In the topography mode the resolution of received image depends on such factors as the number of points of the scanning field (digital resolution), scanning frequency (the so-called scan rate), values of I Gain, P Gain and Z Servo Gain, the range of motion of vertical and horizontal piezo scanners, and, of course, the scan size, as well as the parameters of probe and cantilever. As a rule, work on AFM always starts with large scanning fields, and when it is necessary to see the fine structure of the surface, that is, to obtain an image of area with a maximum resolution, one moves to small scanning fields. In this case it is necessary to change the scan options: to reduce the scan rate, to change the digital resolution and to reduce the range of motion of piezo scanner to the lowest possible. When studying the received image of surface topography, in particular surface profile, one should understand, that the received image is the result of integration and averaging of many system parameters and only to a certain approximation it reflects the actual structure of the surface. Perfection of scanning technology and probe quality with each year allows us to come closer to obtaining the real surface topography, but no matter how close is the approaching, anyway the result of measurement will have some interpolations and errors. However, in most of the problems, solved with AFM, there is no need in finding the surface topography with atomic resolution.

In the case of current measurements, 100% accuracy of the result is also unattainable, both because of the above mentioned reasons (integrating and averaging) and because of coarser circumstances. As it was mentioned before, the currents flowing through the probe tip are very small due to the smallness of the tip area. Therefore, for their registration it is necessary to use either a built-in AFM current amplifier or external amplifiers. All these amplifiers allow one to increase the value of current flowing through the probe tip in $10^n$ times, where $n = 3...11$. The principle of choosing the order of current amplification is very simple: initially the minimum order of current amplification $n = 3$, which is duplicated in the software settings, is selected, then the sample is installed in the microscope, the desired voltage, set point, scan rate and other scan parameters are set and an approach of probe to the sample surface is made. After the approach is finished it is necessary to look at the current trace line in the corresponding window. If there is no signal other than noise, then the selected signal gain is insufficient. After this the next value of $n$ (i.e. $n = 4$) is selected (and duplicated in the software settings) and again one looks at the current trace line window. This continues until there appears a signal in the current trace line window (see Fig. 18).

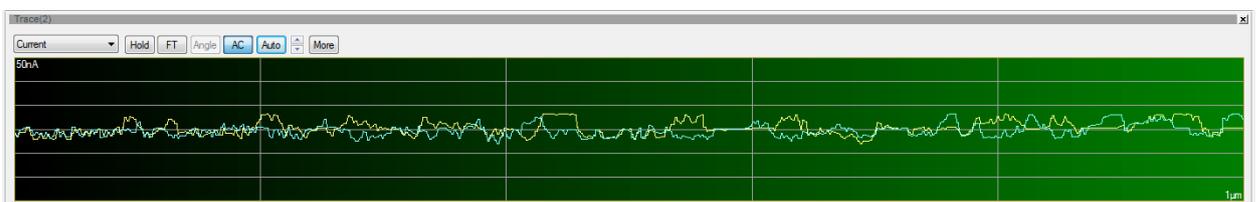

Fig. 18. Screenshots of Current trace lines, depicted at the monitor of AFM PC for the smallest possible value of n.

This means that the chosen value of gain allows one to detect the largest currents along the scan line. But the values of current along the trace line may vary significantly. In order to register not only the current in the separate areas of trace line, but along the entire trace line as well, n should be increased for one more unit. In this case, the following situations are possible:

1. The current profile along the trace line is improved: the signal is detected along the entire length of the trace line (see Fig. 19).

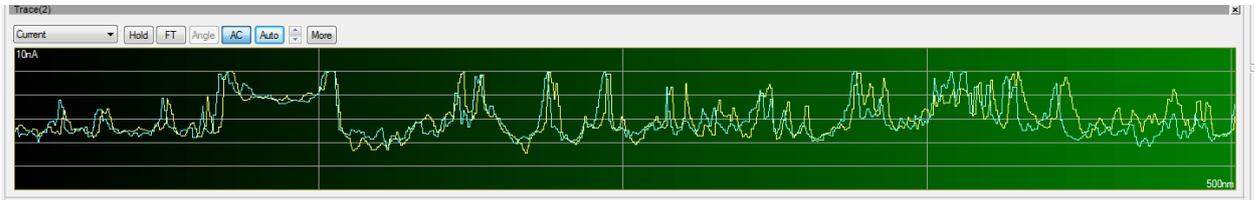

Fig. 19. Screenshots of Current trace lines, depicted at the monitor of AFM PC for the largest possible value of n.

2. The current value exceeds the threshold of the amplifier. Then it is necessary to reduce n by one unit, returning to the previous value of gain (and not to forget to change the value of n in software settings, otherwise the order of current magnitude at the scale of current image will not correspond to reality).

When the user works with samples of the same type, he can immediately set the necessary value of signal gain, that he had chosen before.

Sometimes it happens that in some areas of scanning field there is current, and in the other not. As a rule such islet absence/presence of current is connected with defects or dust on the surface (see the right side of Fig. 20).

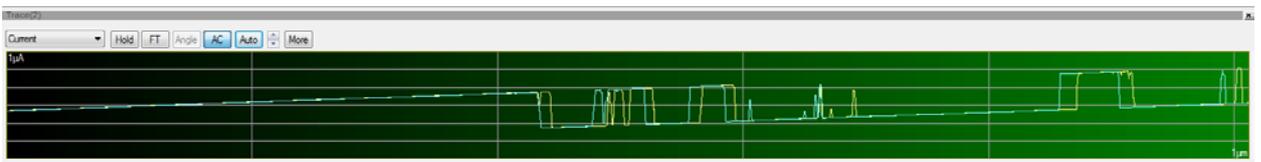

Fig. 20. Screenshots of Current trace lines, depicted at the monitor of AFM PC for the case of defects on sample surface. Sample bias was +1.5V. Low current corresponds to defect areas.

In cases when a large current flows through the defect areas, increasing of n leads to the overload of amplifier and current values, that do not fit in height in the current trace line window, unfold in the opposite direction. It turns out that the current is huge, but is registered only in some places. In the other parts of scanning field there will be only a uniformly colored area, corresponding to the absence of signal. In such cases it is recommended to shift/decrease the scanning field in order to avoid defect areas through which large currents flow and do not allow to detect currents through other areas of scanning field. After this it is necessary to increase the value of n until it will be possible to detect a current on the larger area of scanning field.

As can be seen from the description of current registration procedure, the current image is primarily determined by the signal amplification, set by the user [13]. The user tries to increase the signal to the maximum possible value. However, it should be understood that there are always smaller currents, that flow but are not registered and therefore do not contribute to the resulting current image, simply because for their registration one needs a larger value of gain, that is impossible because of the limited throughput of amplifier. The principle of current registration in I-AFM mode is similar to the principle of imaging in topography mode: at first the user scans a large scanning field with a small value of signal gain, and then he decreases the scanning field size, excluding those areas, through which large currents flow (just like he decreases scan field, eliminating the dust when receiving topography image) and increasing the value of signal amplification (just as he increases the resolution and reduces the scanning frequency and range of motion of piezo scanner in obtaining surface topography).

Thus, the resulting current image is first of all determined by the bandwidth of amplifier, and secondly by the probe parameters and software settings. One should keep in mind that the absence of current in some parts of the current image does not mean that these areas are insulating. This means that either the areas are very narrow and the probe simply do not penetrate them (this needs to be coordinated with the topography image) – then one will get an image similar to that shown in Fig. 8, or (if the topography is a smooth surface), this means that for the selected value of signal gain the user could not register a current through the given areas of scanning field. In order to assert that some areas are insulators, one needs to narrow the scanning field to the dimensions of these areas and to enhance step by step the signal to the maximum possible values of the amplifier. Only in the case if, over the entire range of amplifier, for the selected questionable areas one failed to register any current values, although in other areas of the sample surface, scanned with the same probe, both before and after the study of questionable areas, the current was registered (which means that the probe works properly), it would be possible to assert that the questionable areas are insulating.

Thus, the necessity of using internal/external current amplifier means that the resulting image is built not on the basis of all currents, flowing through each point of the sample surface, but only on those that are registered by the amplifier for the chosen value of gain. I.e., a part of information is inevitably lost. Furthermore, in the investigation of the interfaces of different materials on the sample surface, it may happen, that the selected value of current amplification is not suitable for all substances simultaneously. In this case, one should switch to the internal current amplifier and use a "Log Current" channel, which allows to measure a wider range of currents, transformed through logarithmic circuit. However, this does not always help since in any case, the possibilities of the method are limited by the capabilities of the amplifier.

In view of all listed features of measurement, method of I-AFM in the form, in which it exists today, serves as a qualitative research method that does not allow to determine the actual value of current density due to non-constant area of electrical contact at each scanning point.

### 3. SSRM and SCM modes.

SSRM – is an I-AFM mode, in which scanning surface represents the cross section of the sample (transistor, diode, etc.). The difference lies in the use of a special holder for the vertical cut of the sample. The mode is intended for the study of resistance of layers. If to the cross sections of a diode or a transistor, mounted on a vertical sample holder, we attach an external bias, then switching to the SCM (scanning capacitance microscopy) mode, we will be able to explore the interfaces of semiconductors or metals and semiconductors, in particular, to determine the width of the space charge region (SCR).

When working with sample cross sections there appear a number difficulties. As a rule when a multilayer sample is cut or broken, each layer breaks down differently. As a result, the cleaved surface is not plane, but is inclined with high roughness at the interfaces, that often exceeds the maximum allowable size of roughness for AFM (Z scanner's maximum movable range) – 12 μm [14].

In our case, there was made an attempt to explore the interfaces by means of SCM (to detect the size of SCR when external bias is applied to the sample) and SSRM (to determine the resistance of layers) in the following structures: a-saphire/ZnO:Ga($\sim 1$ μm)/ZnO($\sim 2$ μm)/PdO$_x$/Pd. The total thickness of PdO$_x$/Pd layers was less than 100 nm. The edges of the sample cleavage were sharp and heterogeneous. PdO$_x$/Pd layers were often split off and they were absent at many places of the interface when viewing in an optical microscope. Moreover, when scanning the edge there is always a risk of damaging the probe. In order to protect the probe from damage on the edge of the sample, the sample surface was covered with photoresist. After drying of photoresist the sample was scribed from the sapphire substrate site and split off (see Fig. 21). Fig. 22 shows the SEI image of the end face of our sample.

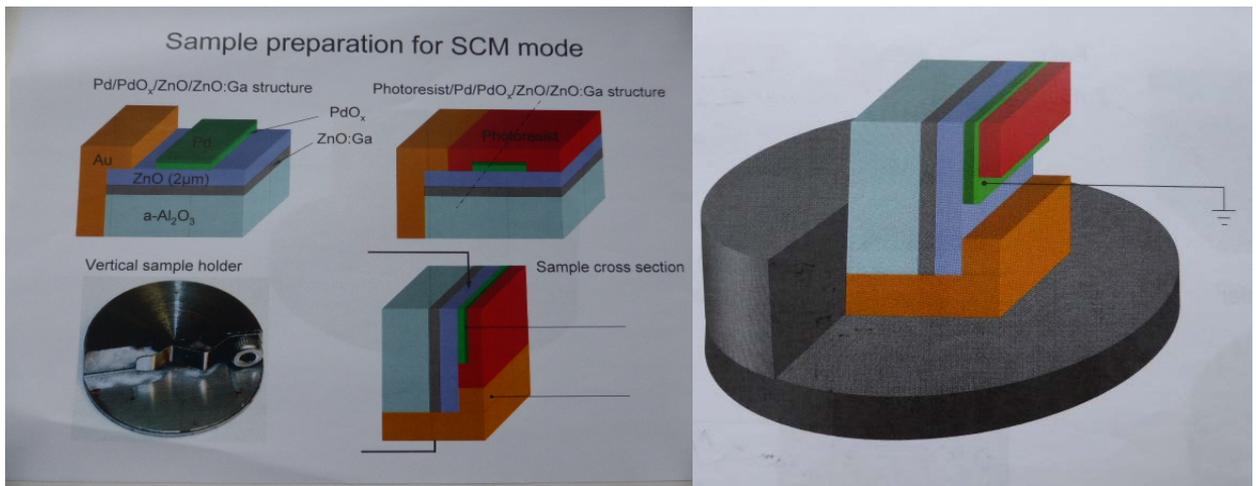

Fig. 21. Schematic representation of Schottky diodes with photoresist on the top, mounted on a vertical sample holder for SSRM and SCM measurements.

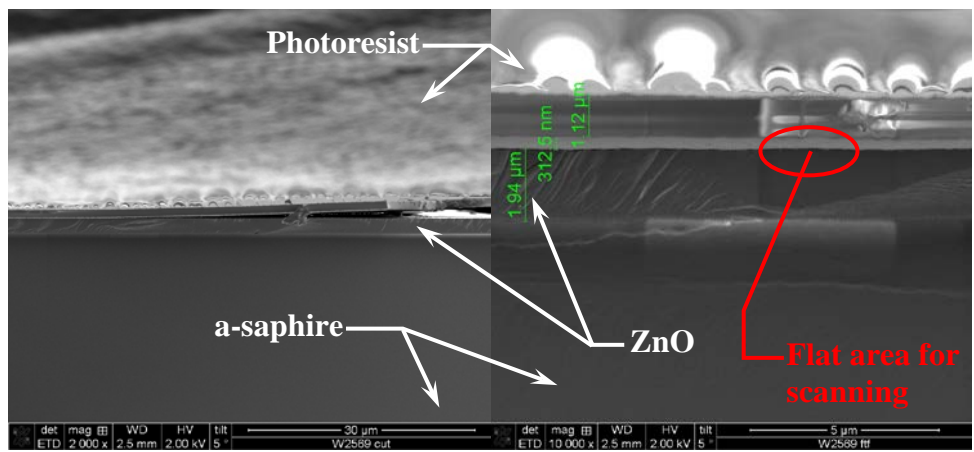

Fig. 22. SEI images of cross section of Schottky diodes with photoresist deposited on the top in order to protect the probe tip from scraping.

As can be seen, the SEM image gives a clear picture of layer boundaries. The photoresist layer is visible from above. It is seen that a part of $PdO_x$/Pd layers was split off (Fig. 22, on the left) during the sample split, and if we scan the sample in a place of layer's split, we will simply break the probe. But there are also homogeneous flat areas suitable for scanning. Their magnified image is shown in Fig. 22 on the right.

However, after the sample was mounted in the vertical sample holder and we tried to approach the probe to the interface, in the optical microscope of SPM (which is used for positioning the probe when it is approached to the surface) we saw only a sapphire (gray on the left) and photoresist (black on the right) (see Fig. 23). The magnification of optical microscope does not allow to see the layer boundaries, as well as the area of layer's split.

Numerous attempts to approach the probe to the visually flat cleavage areas were unsuccessful: even at Z scanner's maximum movable range the cantilever terribly twisted or bent – it clunged to the layer boundaries.

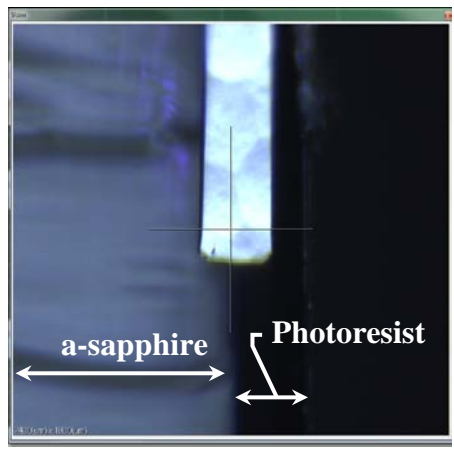

Fig. 23. Screenshot of View window at the monitor of AFM PC when the tip is approached to the sample cross section. To the left from the cantilever is a diode, to the right (dark black area) – the photoresist, that protects cantilever tip from scraping at the edge of the diode.

To solve this problem we tried to polish the end face of the sample with an ion beam in SEM [15]. For this the end face of the sample was initially polished mechanically and then in the SEM. Since the photoresist is rather soft, the mechanical polishing of the end face leads to the rapid abrasion of photoresist. That is why we decided to abandon the use of the photoresist. To avoid the breaking of the probe on the edge of the sample during approach, we split the sample, and then glued the two halves of the split so that the metal and metal oxide were inside the end face and the sapphire – outside (see Fig. 24).

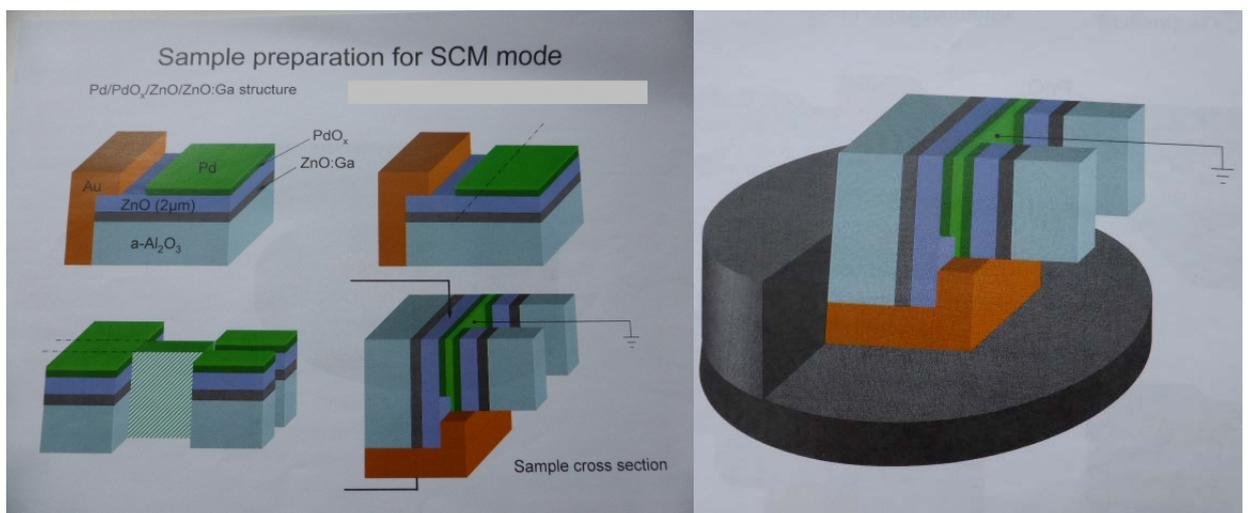

Fig. 24. Schematic representation of Schottky diodes. The diode was cut into 2 parts along the dotted line. These 2 parts were mounted face to face in order to protect the cantilever tip from scraping at the edge of the diode at SSRM and SCM measurements.

The experience showed that the use of vertical sample holder turned out to be undesirable for our samples. It was much easier to fix the sample on a magnetic sample holder using a silver paste (see Fig. 25) in a similar manner to that described for I-AFM (see Fig. 5).

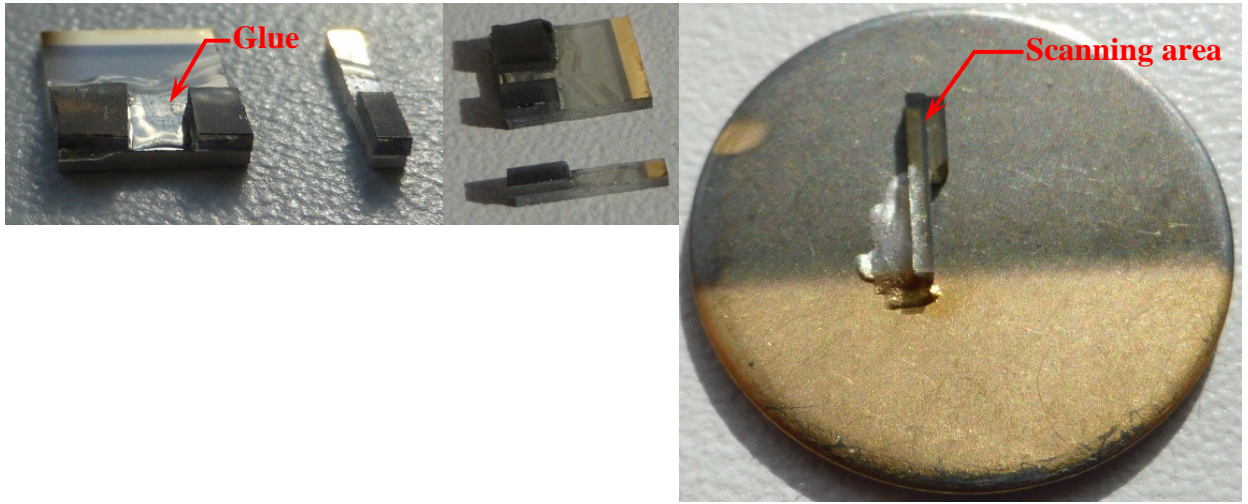

Fig. 25. Images of glued face to face 2 parts of Schottky diode and of mounted on a magnetic sample holder cross section of Schottky diode for SSRM measurements.

The result of mechanical polishing of such cross section is presented in Fig. 26 on the left, and of ion beam polishing in SEM – on the right. Thus, the layers we are interested in, are located on the left and right of the central layer of glue.

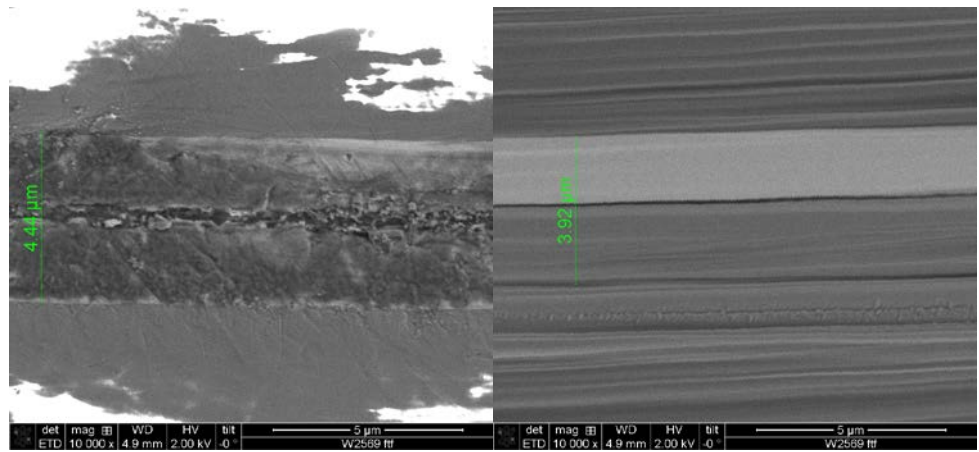

Fig. 26. SEI images of cross section of glued face to face 2 parts of Schottky diode after mechanical (left) and after ion-beam (right) polishing.

At the same time the layer of glue in the middle of cross section is not visible in an optical microscope of AFM. Therefore first of all, using high range modes of XY and Z scanners, we scanned the interface (see Fig. 27), and then we planned to choose flat areas on either side of it and to scan them using low range modes of XY and Z scanners.

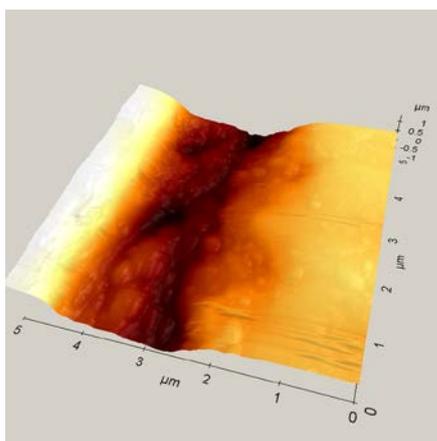

Fig. 27. Topography I-AFM image of the interface between glued face to face 2 parts of Schottky diode.

This raised another problem, this time connected with software. When the range mode of XY scanner was switched from high (up to 100 µm) to low (up to 10 µm), there appeared a displacement relative to the point of the sample surface, where the probe was located. This is due to the fact that high and low range modes of XY scanner have the same origin of coordinates, which is located at the bottom left corner of the scanning field. Therefore, when working in a high range mode of XY scanner, the user switches to a low range mode, then piezo scanner moves to the bottom left corner of the large scanning field, and from this corner begins the scanning field for a low range mode of XY scanner. Thus, when using version XEP 1.8.0 of data acquisition program for XE series SPM, the scanning of vertical cross sections of samples will be almost impossible, because as soon as the user finds the area he is interested in and wants to switch to a low range mode of XY scanner to increase the resolution, piezo scanner moves the sample under the probe to another point. The offset of X and Y scanner coordinates does not help in this case.

## 4. I/V Spectroscopy mode
### The idea of the method

I/V Spectroscopy mode is used to obtain current-voltage characteristics (CVC) in the selected point on the sample surface. As a rule this mode is used in addition to I-AFM mode, when, after getting a current image, it is necessary to obtain and compare CVCs for heterogeneous areas on the current image. In this case, all settings of current amplifier remain unchanged, and the user simply switches from I-AFM to I/V Spectroscopy mode. In I/V Spectroscopy mode, the trace control window is replaced by I/V curve plot window (see Fig. 28). I/V Spectroscopy mode can also be used independently, but this is equivalent to the blind spectroscopy (such use will allow, for example, to receive spectroscopy of defect or dust, etc.). Therefore, without the prior acquisition of topography image, and, what is better, – both topography and current images, the use of I/V Spectroscopy mode would not be rational.

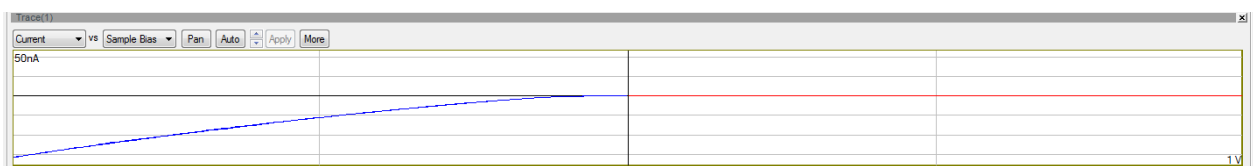

Fig. 28. Screenshots of I/V curves, that appear in the trace control window instead of trace lines, depicted at the monitor of AFM PC when one switches to I/V Spectroscopy mode.

When switching to I/V Spectroscopy mode, it is necessary to specify, which of the two images (a topography or a current image) should be taken as a reference image, on which the user selects the coordinate locations, where he will obtain I/V curves.

Therefore, when the user wants to obtain CVC at some particular point on the sample surface, he instinctively tends to choose topography image as reference image. However, it is necessary to keep in mind that CVC will be obtained not at the point, where the probe tip was sent after selecting of coordinates, but at the place, where the electrical contact of the probe and the surface will take place. This means that when the probe is directed at the point with given coordinates, it is positioned above this point, but, after approaching the probe to the surface, the CVC can be obtained as at this point (if the electrical contact is strictly between the probe tip and the surface), and at some point or area around the selected point (if the contact is between the lateral surface of the probe and the sample, or if the tip coating is worn away). Therefore, one should always compare the point he is interested in, on the topography and on the current image. It may happen that no signal corresponds to this point on the current image, which means that it has no sense to receive CVC at this point. It may also happen (and rather often) that forward and backward topography images are identical, while forward and backward current images are displaced from each other (see Fig. 13). The explanation of this artifact was given above and is connected with the difference in principles of the construction of topography and current images. In addition, it indicates the possible problems with the probe tip (dust, asymmetry, worn out metal coating), and, hence, when obtaining CVCs the user probably will not get them in points of interest.

But it is not the only problem, that arises when operating in I/V Spectroscopy mode. Much more significant problem is that the user in principle can not return to the selected point on the reference image, obtained in I-AFM mode. There are two reasons for this:

1) the thermal expansion of piezoceramics during the operation;

Since electric current flows through piezoceramics, it always leads to its heating. This means that, after receiving each next surface scan, the piezoceramics becomes warmer. When the user sets the coordinates and expects that the probe will come back to a point with these coordinates for obtaining the CVC in this point, then, strictly speaking, the probe will not get to this point as the volume of piezoceramics has increased due to heating, and further expansion or contraction of piezoceramics when applying a voltage will occur relative to its current volume, not the volume which it had at the moment of scanning the point, the user is interested in. This means that contraction of piezoceramics occurs to a value, lower than necessary. Therefore, the probe moves in the region around a predetermined point. Moreover, the longer the time of imaging in I-AFM mode, the greater the discrepancy between the actual position of the probe and the coordinates on the reference image in which it was directed.

2) the very nature of piezoceramics

Piezoceramics, used in the AFM, is made of ferromagnetic materials. And this means that it is inherent in hysteresis effect.

The movement of the sample under the probe during the scanning process can be detected in several ways. The first, and most visible way, is the obtaining of repeat image either by selecting a repeat scan option, or by rescanning the same area at constant settings. One can see that the topography of the first and second scan differ not only in quality, but also in the coordinates of defect location. It is seen especially clear when operating at low range mode of XY scanner with large time interval between the first and second scans (see Fig. 29).

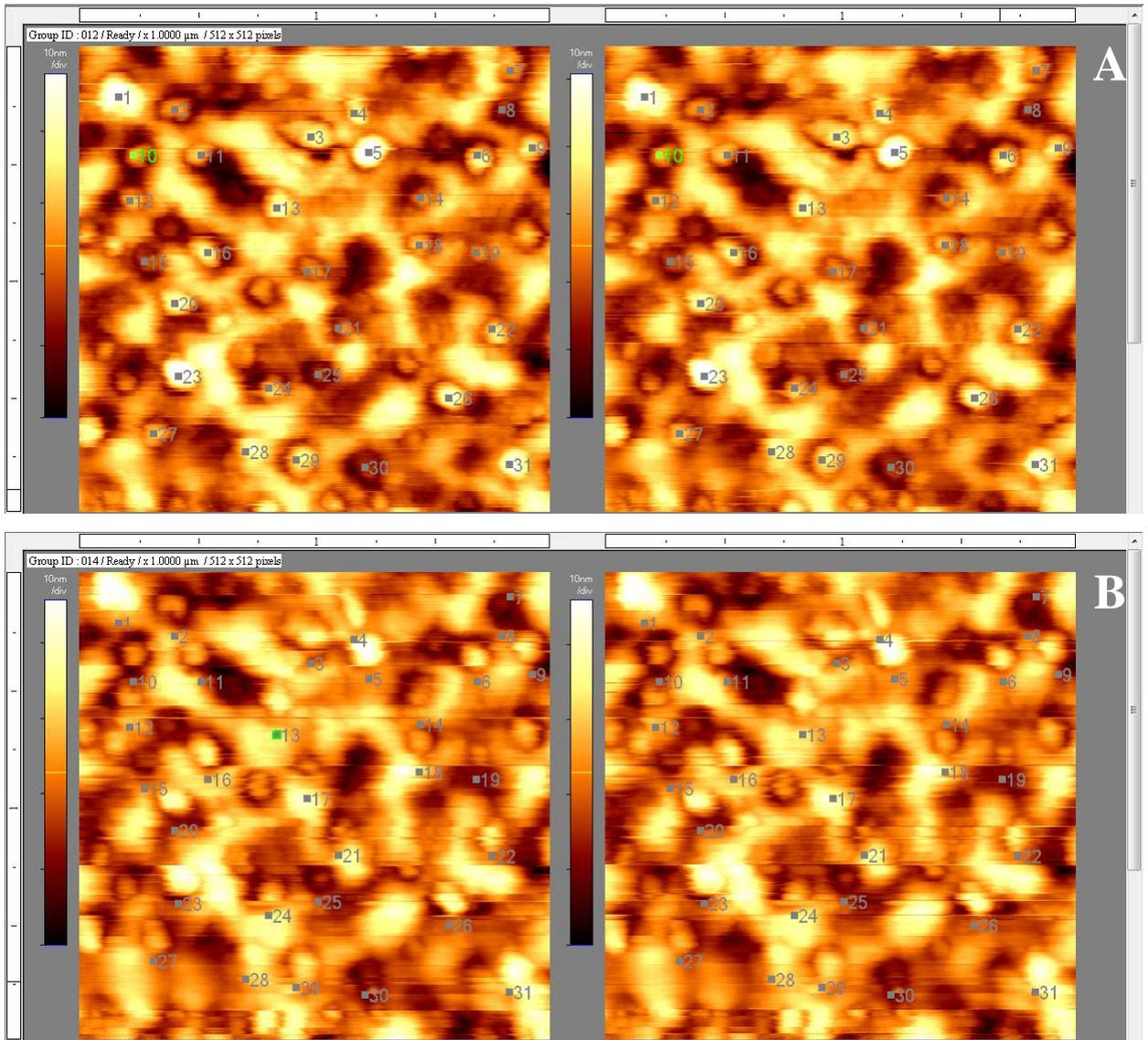

Fig. 29. Screenshots of taken in I-AFM mode topography images with selected points for I/V Spectroscopy measurements. Image A was taken 20 minutes earlier than image B. After image A was obtained in I-AFM mode, we switched to I/V Spectroscopy mode and chose 31 points on topography image for I/V Spectroscopy measurements. After these measurements were made, we switched back to I-AFM mode and, without any changes in settings, we obtained topography image again. Then we switched to I/V Spectroscopy mode and used the coordinates of 31 points for I/V measurements, that remained from previous measurement. One can easily see that in 20 minutes the sample has shifted under the probe. It means that when SPM obtaines I/V characteristics in the first point, these characteristics correspond to this point, but when in about 20 minutes it obtaines I/V characteristics in the last 31st point, this point would have been already shifted and the obtained data will not correspond to the selected point.

The second way – is the use of I/V Spectroscopy mode. We tried to choose the best time for acquiring the CVC at constant number of points – the shortest measurement time, when there was no noise on CVC. We started with a minimum time and gradually increased it. When we compared the CVCs obtained in one and the same point over time, we saw that they are qualitatively different (see Fig. 30). That is, when we were choosing the time for obtaining the CVC, the sample managed to shift under the probe into one of the neighboring points.

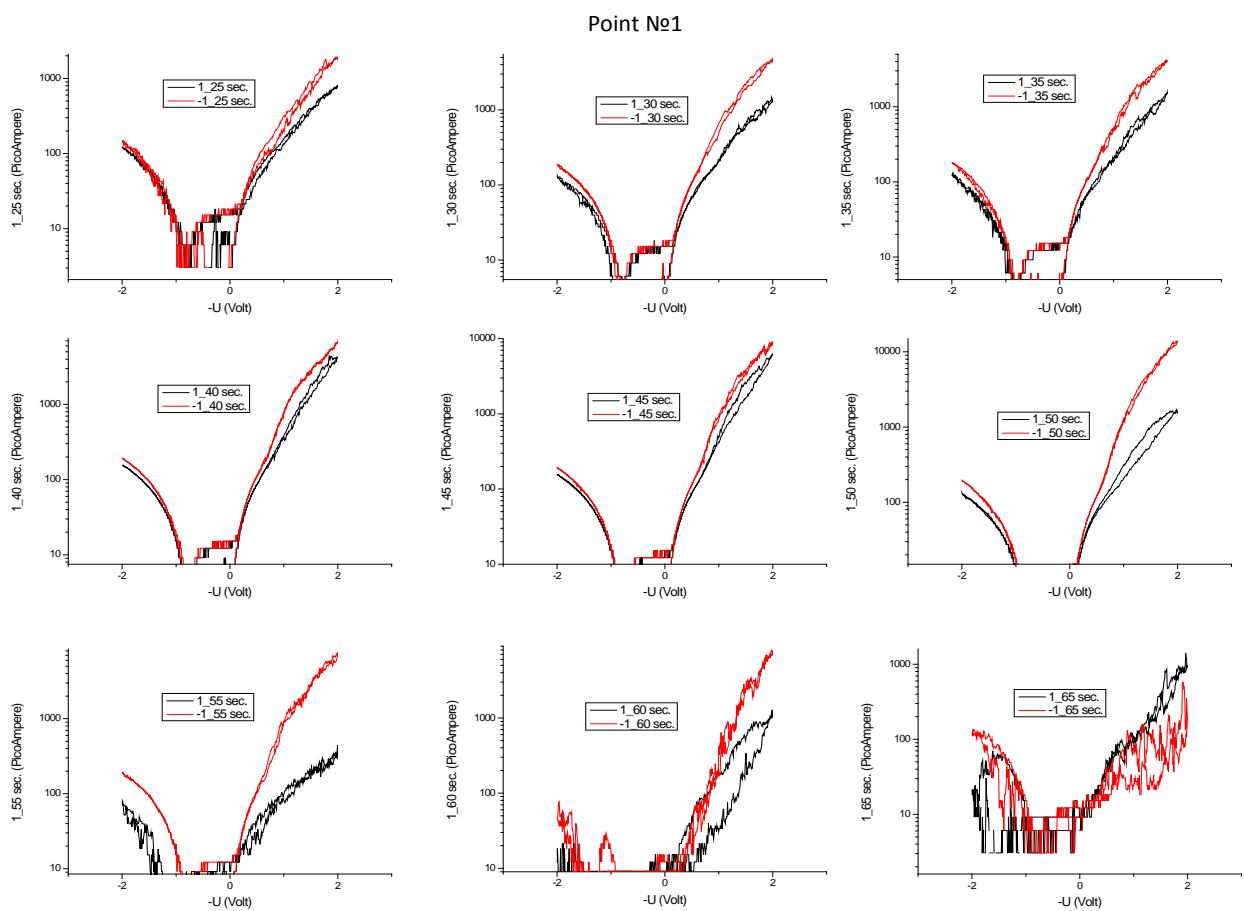

Fig. 30. Results of I/V measurements taken in one and the same point.

We tried to choose the optimal measuring time for CVCs obtained in the voltage range from -2V to 2V for 512 points in one trace. We increased the measuring time from 25 to 70 seconds. Then we decreased the measuring time back from 70 to 25 seconds. Black colour corresponds to forward direction (when measuring time was gradually increased) and red colour – to backward direction, i.e. there is a significant time interval between black and red curves. One can see that 30 seconds is a good time period for I/V measurements. Besides, one can see that curves with different measuring time differ a lot.

From the foregoing it follows that the use of I/V Spectroscopy mode for obtaining the CVCs at the points on topography image, the user is interested in, is not possible.

The idea of the method is promising, but it requires serious improvements:

1) the improvement of the probes (reduction of tip radius and increase in the symmetry and wear resistance);
2) the exclusion of the temperature drift of piezoceramics by combining it with the thermostat;
3) the elimination of drift associated with hysteresis, by more thorough programming.

In case the method works perfectly, it can be successfully used for electrophysical studies of nanostructures. But in the form, in which it is presented today, when the tip curvature radius is up to 150 nm (see Fig. 31), when there is no certainty at what point the probe interacts with the surface (due to the drift of piezoceramics) and what is the area of electrical contact in this place, the investigation of nanostructures is very doubtful.

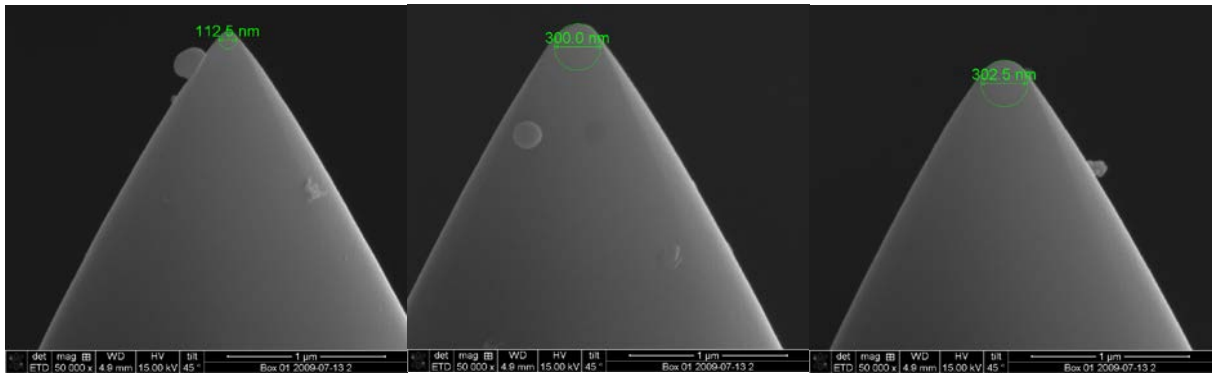

Fig. 31. SEM images of NSC36Ti-Pt probes (A, B, C). One can see that real tip curvature radius exceeds the maximum possible tip curvature radius stated by manufacturer and equal to 40 nm.

Today the method of I/V Spectroscopy can be used only for the average analysis of a homogeneous surface of small area. For example, for obtaining CVCs of Schottky diodes of small dimensions, when Schottky contacts have a size of ~ 500 nm, and are covered with metal capping layer for potential equalization (see Fig. 32).

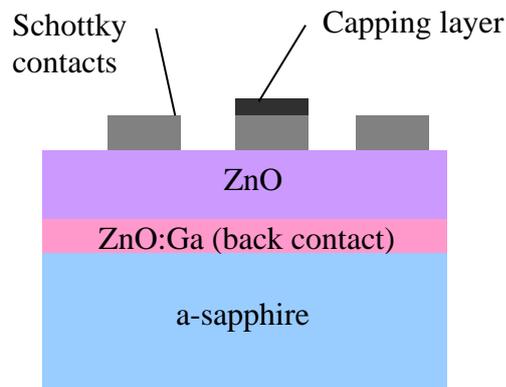

Fig. 32. Schematic representation of cross section of Schottky diode with capping layer.

## Conclusion

1. I-AFM mode is recommended for small scanning fields, when changes in the parameters of the probe during the scanning process can be neglected.

2. It is impossible to compare the topography image with points of obtaining the CVCs, since they are shifted due to the temperature drift and hysteresis of piezoceramics.

3. The use of I-AFM, I/V Spectroscopy and especially SSRM and SCM modes will be more effective if they are available for AFM, integrated in SEM. In such a case, firstly, the scanning is carried out not in air but in vacuum, that increases the resolution of the method, secondly, when operating in SSRM mode the user can accurately position the tip above the layer boundary, which is not visible in an optical microscope mounted on XE-150, and thirdly periodically switching on the SEM, the user can monitor the status of the probe tip.

## Acknowledgements

The material presented in the article was obtained during the execution of the grant Webb Erasmus Mundus Action 2 Lot 5 – strand 1 2012-2739/001-001-EM 2 Partnerships.

All experiments presented in the paper were made on the equipment of Semiconductor Physics Group from Institute for Experimental Physics II, University of Leipzig. I thank Gabriele Ramm